\documentstyle[epsfig,twoside,fleqn,espcrc2]{article}

\title{
On the Analytic ``Causal" Model for  the QCD Running Coupling}
\author{D.V. Shirkov\address{Bogoliubov Lab. of Theor. Physics, JINR,
Dubna, 141980, Russia;\\  e-mail:  shirkovd@thsun1.jinr.dubna.su}}

\begin{document}

\begin{abstract}
We discuss the model $\bar{\alpha}_{\rm an}(Q^2)$ recently proposed for
the QCD running coupling $\bar{\alpha}_s(Q^2)$  in the Euclidean domain
on the basis of the ``asymptotic--freedom" expression and on ca\-usality
condition in the form of the $Q^2$-ana\-ly\-ti\-city.

The model contains no  adjustable parameters and obeys the
important features:
(i) Finite ghost-free behavior in the ``low $Q^2$" region and
correspondence with the standard RG-summed UV expressions;%\item~ %
(ii) The universal limiting value $\bar{\alpha}_{\rm an}(0)$
expressed only via group symmetry factors. This value as well as
the $\bar{\alpha}_{\rm an}$ behavior in the whole IR region
$Q^2 \leq \Lambda^2$ turns out to be stable with respect to higher
loop corrections; %\item~
(iii) Coherence between observed $\bar{\alpha}_s(M_{\tau}^2)$
value and integral information on the IR  $\bar{\alpha}_s(Q^2)$ behavior
 extracted from jet physics.
\end{abstract}

\maketitle

\section{INTRODUCTION}

This presentation is devoted to the review and discussion of a new
{\it analytized} model expression $\bar{\alpha}_{\rm an}(Q^2)$ for
the QCD running coupling recently devised~\cite{jinr} by combining
the RG-summed expression $\bar{\alpha}_s(Q^2)$ with the demand of
analyticity, that is causality, in the $Q^2$ complex plane.
This procedure "cures" the IR ghost-pole trouble by an additional
contribution that is non-analytic in the coupling constant and at the
same time preserves the asymptotic freedom property and
correspondence with perturbation theory in the UV . \par

    The ``analytization procedure'' elaborated in the mid-fifties (see
 Ref.\cite{trio}) consists of three steps:   \par
--  Find an explicit expression for $\bar{\alpha}_{RG}(Q^2)$ in the
Euclidean region $Q^2>0$ by standard RG improvement of a perturbative
input. \par
--   Perform the straightforward analytical continuation of this
expression into the Min\-kow\-skian region ${\rm Re}\, Q^2<0 ,{\rm Im}\,
Q^2=-\epsilon$. Calculate its imaginary part and define the spectral
density by $\rho_{RG} (\sigma ,\alpha)=
{\rm Im}\bar{\alpha}_{RG}(-\sigma -i\epsilon ,\alpha)$. \par
--  Using the spectral representation of the K\"allen--Lehmann
type %see Eq.(\ref{spectral}) below,
with $\rho_{RG}$ in the
integrand, define an ``analytically-improved'' running
coupling $\bar{\alpha}_{\rm an}(Q^2)$ in the Euclidean region.
\smallskip

Being applied to $\bar{\alpha}(Q^2)$ in the one- and two-loop QED
\{or QCD\} case, this procedure produces (see Ref.~\cite{trio}
\{or ~\cite{jinr,prl97}\}) an expression
 $\bar{\alpha}_{\rm an}(Q^2)$ with the properties:\par
\smallskip

(a) it has no ghost pole,                    \par
(b) in the complex  $\alpha$-plane at the point $\alpha=0$ it
   possesses an essential singularity  $\sim \exp(-1/\alpha \beta_1)$,
   with $\beta_1$, the one-loop beta-function coefficient, \par
(c) in the vicinity of this singularity for real and positive
   $\alpha$ it admits a power expansion that coincides with the
   perturbation input, \par
(d) it has the finite UV \{or IR\} limit $ 4\pi/\beta_1 $ that {\it
does not depend on the experimental value} $\alpha\simeq 1/137$
\{or $\Lambda\}$.

 In the one-loop QCD case this expression is of the form
\begin{equation} \label{a1} \bar{a}^{(1)}_{\rm an}(Q^2) =
\frac{1}{\ln Q^2/\Lambda^2}\,+\,\frac{\Lambda^2}{\Lambda^2-Q^2}
\end{equation}
with $\bar{a}=\beta_1\bar{\alpha}_s/4\pi$ and
$\beta_1=11-(2/3)n_f$.

    The same procedure being applied to the two-loop  case
yields~\cite{trio,jinr,prl97} a more complicated expression with the
same essential features (see, e.g., Eqs. (\ref{rho2}) - (\ref{spectral})
below).   % As it has been demonstrated (see, also Ref.~\cite{prl97})

\section{THE MODEL FOR QCD RUNNING COUPLING}

\mbox{The~~``analytic'' coupling constant, Eq.~(\ref{a1}),} instead of
ghost pole has a weak singularity at $Q^2=0$ and its IR limit
$\bar{\alpha}_s^{(1)}(0)=4\pi/\beta_1 $ depends only on group factors.
Numerically, for $n_f=3$,we have $\bar{\alpha}_{\rm an}^{(1)}(0)=
4\pi/9\simeq 1.4$.

 Note, to relate $\Lambda$, the QCD scale parameter, to $a(\mu^2)$,
in our case we have to change its usual one-loop definition for
$\Lambda^2=\mu^2\exp[-\phi(a(\mu^2))]$
with the function $\phi$ defined by the transcedental relation
\begin{equation} \label{trans}
a=1/{\phi(a)}-1/[\exp (\phi (a))-1]~\,
\end{equation}
which is consistent with the symmetry property
\begin{equation} \label{symm}
\ \phi (a) = - \phi (1-a) \, .
\end{equation}

The corresponding beta-function
\begin{equation}
\beta (a) = - \frac{1}{\phi^2(a)} + \frac{\exp{\phi(a)}}
{\left[\exp{\phi(a)} -1\right]^2}
\end{equation}
is also symmetric $\beta (a) = \beta (1-a)$ %due to Eq.(\ref{symm})
and obeys the second-order zero at $a=1$ -- see Fig.~1.
%
%%%%%%%%%%%%%%%%%%%%%%%%%% FIGURE 1 %%%%%%%%%%%%%%%%%%%%%%%%
%
 \begin{figure}[ht]
 \centerline{
 \epsfig{file=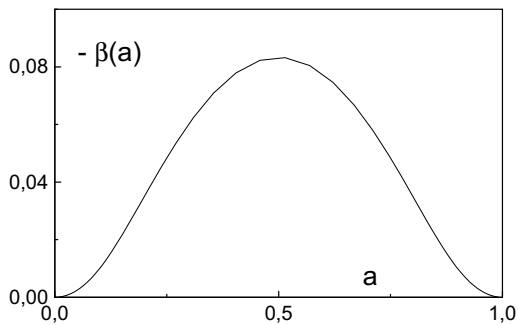,width=7.8cm}
    }
 \vspace{-1.1cm}
 \caption{ One-loop $ \beta(a)$ function.}
 \label{beta_aa}   \end{figure}

Usually, in perturbative QFT practice, we are accustomed to the
idea that theory supplies us with a set of possible curves for
$\bar{\alpha}_s(Q)$ and one has to fix the ``physical one'' by
comparing with experiment. Here, Eq.~(\ref{a1}) describes a family of
such curves for $\bar{\alpha}_{\rm an}(Q^2)$ forming a bundle with
the same common limit at $Q^2=0$.

For the two-loop case, we start with the running coupling written
down in the form
\begin{equation} \label{af2}
\bar{a}^{(2)}_{RG}(Q^2)=\frac{1}{l+b \ln (1+l/b)}~ ,
\quad l=\ln\frac{Q^2}{\Lambda^2}\, ,
\end{equation}
with $b=\beta_2/(\beta_1)^2$ and $\beta_2=102-38/3n_f$. This
expression corresponds to the result of exact integration of the
two-loop differential RG equation explicitly resolved by iteration.
It generates a popular two-loop formula with the $\ln l/l^2$ term.
For the spectral density, we have
\begin{equation}        \label{rho2}
 \rho^{(2)}_{RG} (L)\,=\,\frac{I(L)}{R^2(L)\,+\,I^2(L)}\, ,
\quad L=\ln\frac{\sigma}{\Lambda^2}
\end{equation}
with
\begin{equation}               \label{rl}
R(L)=L+b\ln \sqrt{(1+L/b)^2+ (\pi/b)^2}~,
\end{equation}
\begin{equation}               \label{il}
 I(L)=\pi+b\arccos\frac{b+L}
{\sqrt{\left(b+L\right)^2+\pi^2}}~.
\end{equation}
Now, to obtain $\bar{a}^{(2)}_{\rm an}(Q^2)$, one has to substitute
Eq.~(\ref{rho2}) into the r.h.s. of
\begin{equation}  \label{spectral}
\bar{a}_{\rm an}(Q^2)= \frac{1}{\pi} \int_0^\infty
d\sigma \frac{\rho (\sigma ,a)}{\sigma+Q^2- i \epsilon}\, .
\end{equation}
The one-loop result, Eq.(\ref{a1}), follows from Eqs.~(\ref{rho2}) -
(\ref{spectral}) at $b=0$. However, in the two-loop case, the
integral expression thus obtained is too complicated for being
presented in an explicit form. For a quantitative discussion one has
to use numerical calculation.   \par
   Nevertheless, for a particular value at $Q^2=0$ we can make
two important statements. First, the IR limiting coupling value
$\bar{\alpha}_{\rm an}(0)$ does not depend on the scale parameter
$\Lambda$. Second, this value turns out to be defined by the one-loop
approximation, i.e., its two-loop value coincides with the one-loop
one (for detail see ref.~\cite{prl97}). \par
   Thus, the $\bar{\alpha}_{\rm an}(0)$ value, due to the RG
invariance, is independent of $\Lambda$ and, due to the analytic
properties,  of higher corrections. This means that the analyticity
stabilizes the running coupling behavior in the IR, makes it
universal. \par
     Moreover, the whole shape of
the $\bar{\alpha}_{\rm an}(Q^2)$ evolution turns out to be quite
stable with respect to higher corrections. The point is that the
universality of $\bar{\alpha}_{\rm an}(0)$  gives rise to stability
of the ${\bar{\alpha}_{\rm an}^{(\ell)}(Q^2)}$ behavior with
respect to higher correction in the whole IR region (at $Q^2
\simeq \Lambda^2$). On the other hand, stability in the UV domain is
reflection of the property of asymptotic freedom. As a result, our
analytical model obeys approximate ``higher loops stability" in the
whole Euclidean region. Numerical calculation (performed in the
$\overline{\rm MS}$ scheme for $\ell = 1,2$ and 3 cases with
$n_f = 3$) reveals that ${\bar{\alpha}_{\rm an}^{(2)}(Q^2)}$ differs
from ${\bar{\alpha}_{\rm an}^{(1)}(Q^2)}$ by less than  10\% and
${\bar{\alpha}_{\rm an}^{(3)}(Q^2)}$ from
${\bar{\alpha}_{\rm an}^{(2)}(Q^2)}$ within the 1\% limit, as it is
shown in Fig.~2.
%
%%%%%%%%%%%%%%%%%%%%%%%%%% FIGURE 2 %%%%%%%%%%%%%%%%%%%%%%%%
%
 \begin{figure}[ht]
 \centerline{
 \epsfig{file=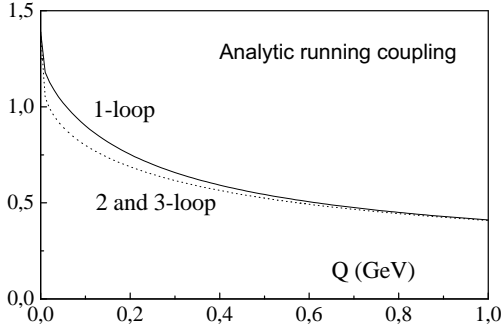,width=7.8cm}
    }
 \vspace{-1.1cm}
 \caption{ ``Higher loop stability" of the analytic solution.
 The normalization point is on the $\tau$ lepton scale:
 $\bar{\alpha}_s(M_{\tau}^2)=0.34$.
 }
 \label{beta_a}  \end{figure}

The idea that QCD running coupling can be frozen or finite at small
momenta was considered in many papers (see, e.g., the discussion in
Ref.~\cite{M-Stev} and interesting theoretical model in Refs.
\cite{sim}). There seems to be experimental evidence in favor of such
behavior. As an appropriate object, for comparison with our
construction, we use the average
\begin{equation}   \label{aq}
A(Q)\,=\,\frac{1}{Q}\,\int_0^Q\,dk\,{\bar{\alpha}_s}(k^2)
\end{equation}
\noindent
that people manage to extract from jet physics data. Empirically, it
has been claimed~\cite{Dok-Webb},~\cite{d-k-t} that this integral at
$Q\simeq 1\div 2$ GeV turns out to be a fit-invariant quantity with
the estimate: $A(2\,\,{\rm GeV})=0.52 \pm 0.10$. Numerical results
for it obtained by the substitution $\bar{\alpha}^{(2)}_{\rm an}$
into Eq.~(\ref{aq}) with $\Lambda_{\rm an} = 710 \pm 110 $ MeV,
corresponding to $\bar{\alpha}_{\rm an}(M_{\tau}^2) = 0.36 \pm 0.02$,
 are summarized in the Table.

\begin{center}
\begin{tabular}{|c|c|c|c|}  \hline
$\bar{\alpha}_{\rm an}(M_{\tau}^2)$ &0.34  &0.36  &0.38 \\ \hline
~$\Lambda_{an}^{(2)}({\rm MeV})$~&~610~ &~710~&~820~
\\ \hline
~$A^{(2)}(2\,{\rm GeV})$~&~0.48~&~0.50~&~0.52~
\\ \hline
\end{tabular}
\end{center}

Note that a nonperturbative contribution, like the second term in
the r.h.s. of Eq.~(\ref{a1}), reveals itself even at moderate $Q$
values by ``slowing down" the velocity of $\bar{\alpha}_s(Q)$
evolution. For instance, at $Q=3\,{\rm GeV}$ it contributes about 4\%,
which  could be essential for the resolution of the ``discrepancy"
between ``low-$Q$" and direct $Z_0$  data for $\bar{\alpha}_s(M_Z)$.

     As far as we have no explicit expression for the
${\bar{\alpha}_{\rm an}(Q^2)}$ in the two-loop case, for a qualitative
discussion we can use an approximate formula proposed in Ref.
\cite{prl97} which can be written in the form of Eq.~(\ref{a1}) with
the substitution
\begin{equation} \label{a2expl}
\frac{Q^2}{\Lambda^2} \to  \\ \exp\left[\ln\,\frac{Q^2}{\Lambda^2}+
b\ln\sqrt{\ln^2\,\frac{Q^2}{\Lambda^2}+4\pi^2}\right].
\end{equation}

With appropriate redefinition of $\Lambda$, the accuracy of this
approximation, for $Q\geq \Lambda$, is no less than 5\%. At the same
time, it produces only a 3\% error in the $A(2\,{\rm GeV})$ value.

\section{DISCUSSION}    % \vspace{2mm}

It is important to discuss the possibility of using
${\bar{\alpha}_{\rm an}(Q^2)}$ in multi-prong QFT objects
$\Gamma_{(k)}(s_1,...,s_k)$ and, particularly, in observables
${\cal M}(... Q^2_i, ... , q^2_j ...)$ with some arguments
$q^2_j= - Q^2_j > 0$ being time-like (Minkowskian) and some others
fixed on the mass shell. \par
   Here, we have in mind a few different items:
\begin{enumerate}
\item The possibility of using RG for $k$-prong vertices
$\Gamma_{(k)}(s_1,...,s_k) \, , k\geq 3$;
\item The technology of using ${\bar{\alpha}_{\rm an}(Q^2)}$,
 originally defined for $Q^2_i>0$, in observables \\
${\cal M}(... Q^2_i, ... , q^2_j ...)$ with time-like arguments;
\item
  The expediency  of the ${\bar{\alpha}_{\rm an}(Q^2)}$
continua\-tion into the Min\-kow\-skian region $Q^2<0$;
\item  The scheme dependence of analytic running coupling. Relation
to divergencies.
\end{enumerate}

   Our preliminary comments on these issues are:

  I. As it is well known from the old investigations, the use of
RG, rigorously deduced from Dyson renormalization
transformations (with finite real counterterm coefficients $z_i$),
is justified only in the Euclidean domain and involves a simultaneous
change of a scale for \underline{all} arguments of $\Gamma_{(k)}(s_j)$,
i.e., $s_j \to s'_j=ts_j$. This restricts the possibility of UV analysis
by the so-called non-exceptional momenta.

II. Nethertheless, in some special cases it turns out to be possible
to apply the RG technique to analyse multi-prong object  by involving
some additional means like spectral representations~\cite{ilya},
light-cone expansion or amplitude factorization~\cite{efr}. \par
   In any of above-mentioned cases the RG procedure results in
solving group equations for real functions, like spectral densities
and light-cone expansion coefficients. This solution involves only real
running coupling ${\bar{\alpha}_{\rm an}(Q^2)}$ values for Euclidean
arguments. Any observable ${\cal M}$ with time-like value of kinematic
invariant $q^2$ should be treated separately with an adequate procedure
of analytic continuation of the observable under consideration. \par
  III. Due to the last reason, the analytic continuation of the running
coupling itself, e.g., discussing of ${\bar{\alpha}_{\rm an}(Q^2)}$
properties in the Min\-kow\-skian region $Q^2<0$, in our opinion, has
no direct physical meaning. \par
  IV. The last but not least is the property of the scheme dependence
of the model discussed. Formally, in our final expressions, Eqs.
(\ref{rho2})--(\ref{spectral}), there is no room for such dependence.
This is related to the absence of UV infinities with their subtraction
and renormalization ambiguities. \par
    The analytical model has an important property. By construction
it is {\it free from UV divergen\-cies}\/. The log squared in the
spectral function denominator (\ref{rho2}) %,(\ref{rl}), (\ref{il})
provides us with convergence of
non-subtracted spectral representation (\ref{spectral}). At the very
end, it contains only one parameter $\Lambda$ that has to be defined
from experiment. However, this needs an adequate procedure (mentioned
above in the comment II) of analytization with possible liberating of
infinities by contribution non-analytic in $\alpha$ for the observable
confronted with data. \\

   To summarize, it can be said that to get more satisfactory answers
and, correspondingly, more complete understanding, it is necessary to
continue investigation of all four issues.

\section*{Acknowledgements} %\vspace{3mm}

The author would like to thank  A.L.~Kataev, E. de Rafael and I.L.
Solovtsov for useful discussions. Partial support by INTAS 93-1180 and
RFBR 96-15-96030 grants is gratefully acknowledged.

\section*{Discussions}
\smallskip %\vspace{3mm}

\noindent{\bf K.Chetyrkin} \\
\noindent {\it What do you think about an experimental testing of
different predictions for the IR behavior of $\bar{\alpha}_s(Q)$ ?} %\\
\vspace{2mm}

\noindent {\bf D. Shirkov} \\
\noindent {\it As I showed in the Table, our model nicely correla\-tes
the $\bar{\alpha}_s(M_\tau)$ measured value with the Khoze-Dokshitzer
integral estimate extracted from the jet physics. This can be compared,
e.g., with the $A(2\,\,{\rm GeV})$ value corresponding to the
Badalian-Simonov model}\/~\cite{sim} {\it which is of an order of 0.35.}

 \end{document}